\documentclass[11pt,english,onehalfspacing,liststotoc, toctotoc, headsepline, consistentlayout,]{article}
\usepackage[top=1.75cm, bottom=1.5cm, left = 1.65cm, right = 1.65cm]{geometry} 
\usepackage{caption}
\usepackage{cite}
\usepackage{longtable}
\usepackage{tabularx}
\usepackage{caption}
\usepackage{graphicx}
\usepackage{subcaption}
\usepackage{amsmath}
\usepackage{textcomp}
\usepackage{blindtext}
\usepackage[utf8]{inputenc}
\usepackage[T1]{fontenc}
\usepackage{palatino} 
\usepackage{float}
\usepackage[autostyle=true]{csquotes}
\usepackage{amsmath}
\usepackage{pdfpages}
\usepackage{xcolor ,graphicx}
\usepackage[labelfont=bf,font={it}]{caption}
\usepackage[labelfont=bf,font={it}]{subcaption}
\usepackage{amsmath}
\usepackage{wasysym}
\usepackage{authblk}
\title{Identical Bands Around the Isobaric Rare Earth Even-Even Nuclei with the Mass Number A = 164}

\author[$\star$]{M. A. Abdelsalam}
\author[$\star$]{H. A. Ghanim}
\author[$\star$]{M. Kotb}
\author[$\star$]{A. M. Khalaf}

\affil[$\star$]{Physics Department, Faculty of Science, Al-Azhar University, Cairo, Egypt}
\affil[\space]{Corresponding author: mahmoudkotb@azhar.edu.eg}

\date{}
\begin{document}
\maketitle

\bibliographystyle{Unsrt}


\begin{abstract}
Eight pairs of rare-earth normally - deformed (ND) nuclei around the isobaric nuclei with A = 164 and have identical values of F-spin, $\pm $ $ F_{0} $ and $ N_{p} $ $ N_{n} $ ($ N_{p} $ and $ N_{n} $ are the number of valence protons and valence neutrons respectively ) have been studied. These pairs of identical bands (IB's) cover 16 mass units and are classified as (i) 3 pairs of nuclei separated by (2p,2n) :($ ^{162}Yb-^{166}Hf $), ($ ^{162}Er-^{166}Yb $), ($ ^{162}Dy-^{166}Er $) (ii) 2 pairs of nuclei separated by (4p,4n): ($ ^{160}Dy-^{168}Yb $), ($ ^{160}Er-^{168}Hf $) (iii) 2 pairs of nuclei separated by (6p,6n): ($ ^{158}Er-^{170}W $) ($ ^{158}Dy-^{170}Hf $)  and (iv) one pair of nuclei separated by (8p,8n): ($ ^{156}Dy-^{172}W $).

We suggested a theoretical collective rotational formula containing three parameters (CRF3) as an extended version of Bohr-Mottelson model to calculate the ground state positive parity excitation energies. Also, the sd-version of the interacting boson model (IBM) has been used to describe the nuclear shapes by using the intrinsic coherent-state. The optimized models parameters for each nucleus are adjusted by using a simulation search program to minimize the root mean square deviation between the theoretical calculation and experimental excitation energies.  The best adopted model parameters of the CRF3 are used to calculate the rotational frequencies $ \hbar\omega $, the kinematic $ J^{(1)} $ and dynamic $ J^{(2)} $ moments of inertia and the evolution of $ J^{(1)} $ and $ J^{(2)} $ with increasing $ \hbar\omega $ are systematically analyzed. A smooth gradual increase in both moments of inertia was seen.

The calculated results agree excellently with the experimental ones which give strong support to the suggested CRF3.

The adopted IBM parameters are used to calculate the potential energy surfaces (PES's) which describe the nuclear deformation. The PES's for our nuclei shows two wells corresponding to prolate and oblate sides which indicate that these nuclei are deformed and have rotational behaviors.

The correlation quantities which identify the IB's are extracted. It is found that the nuclei having $ N_{p} N_{n} / \bigtriangleup $ where $ \bigtriangleup $ is the average pairing gap, exhibit identical excitation energies and energy ratios in their ground state rotational bands.
\end{abstract}


\section{Introduction}

The discovery of rotational bands in adjacent even-even and odd-mass superdeformed (SD) nuclei in which the $ \gamma $-ray transition energies are nearly identical to within a few KeV was an exotic and unexpected phenomenon in nuclear structure physics \cite{001,002,003,004,005}. Since the identical bands (IB's) have essentially identical transition energies, then the associated dynamical moment of inertia are thus identical. Several explanations were put forward \cite{004,005,006,007,008,009,010,0100,011} to understand the origin of IB's phenomenon assuming the occurrence of such IB's to be a specific property of the SD states in nuclei. The explanations of these IB's includes: the Coriolis force, the particle alignment and pairing \cite{012}, the roles of special high-N orbitals of intruder configuration and band crossing\cite{013,014,015,016}, the pseudo-spin in supersymmetry \cite{007,017,018} and the supersymmetry with many-body interactions \cite{019}.

Soon the phenomenon of low-spin identical bands was found in pairs of even-even normal deformed (ND) nuclei \cite{020}, and in neighboring even-even and odd-mass nuclei in rare-earth region where they have similar moments of inertia \cite{021,022}. If was noted that low spin IB's are not limited to nearby nuclei but are widespread and found in pairs of even-even nucleoside as separated by 24 mass unit (like $ ^{156}Dy,^{180}Os $) \cite{023}. Attempts were made to understand the low-spin IB's in terms of some simple systematics of the moments of inertia in the rare-earth region \cite{0241,0242,025,026,027,028} or from several types of consideration \cite{029}.

For the description of normally deformed (ND) bands, some useful models were proposed. Bohr and Mottelson \cite{030} pointed out that, under the adiabatic approximation, the rotational energy of an axially symmetric nucleus may be expanded for $ K = 0$ band as a power series in the I(I+1) term. The expansion for the $ K \neq 0$ band takes the same form, but includes a band head energy and the I(I+1) is replaced by $ \left[I(I+1)-K^{2}\right] $. Another useful models for nuclear rotational spectra are the particle-rotor model (PRM) \cite{031}, the variable moment of inertia (VMI) model \cite{12,21}, the soft rotor model \cite{PhysRevC} and the interacting boson model \cite{n12}.

In the concept of F-spin and its projection \cite{cas} any pairs of conjugate nuclei with the same F-spin and $ F_{0} $ values in any F-multiplet will have the same $ N_{p}N_{n} $ \cite{023,72,1234} where $ N_{p} $ and $ N_{n} $ are respectively the number of valence protons and valence neutrons. The product $ N_{p}N_{n} $ was used in the classification of the changes that occur in nuclear structure \cite{123,040}. It was assumed that \cite{0241,658} the moment and the P-factor depends also on the product $ N_{p}N_{n} $.

The purpose of the present paper is (i) to analyse the excitation energies for even-even normally deformed nuclei in rare earth region in framework of suggested new collective rotational formula (CRF3). (ii) to exhibit the occurrence of IB's in eight pairs of nuclei in rare earth region. (iii) to present the parameters which characterize the appearance of IB's. (iv) use the sd version of interacting boson model (sdIBM) to calculate the potential energy surfaces (PES's).


\section{Outline of the Suggested Collective Rotational Formula with Three Parameters (CRF3)}

Rotational states in normal deformed (ND) nuclei can be characterized by their excitation energies E(I) as a function of spin I, which generally lie low as compared to the single-particle excitation.
In the strong coupling limit, the rotational ground state energy for an axially symmetric even-even nucleus obeys the I(I+1) rule, i.e form bands of levels that fulfill the relation
\begin{align}
	E(I)&=\dfrac{\hbar^{2}}{2J} I(I+1)=\alpha \, \textit{\text{\^{I}}} ^{2}
\end{align}
where $ \alpha $ = $ \hbar^{2}/2J $ and \^{I} = I(I+1)

The relation (1) defines in addition the nuclear moment of inertia J as a constant for an ideal rotor. This simple rotational formula gives deviations from experimental data, So Bohr and Mottelson pointed out that agreement was improved by adding to it a second team to yield
\begin{align}
	E(I) &= \alpha I(I+1) +\beta [I(I+1)]^{2}\nonumber \\
	&=\alpha\,\text{\^{I}}^{2} + \beta\,\text{\^{I}}^{4}\nonumber \\
	E(I)&=\alpha \,\text{\^{I}}^{2} (1+\gamma\, \text{\^{I}}^{2})
\end{align}
where $ \gamma  = \beta / \alpha$

Since the moment of inertia J increases on rotation of the nucleus, the observed deviations from the experiment were still more evident.

According to the variable moment of inertia(VMI) model\cite{12,21}, there is a gradual increase in moment of inertia J with increasing the spin I, so we suggest that the moment inertia J can be written as
\begin{equation}
	J=J(I) =J\,(1\,+\,\sigma\,\text{\^{I}}^{2})
\end{equation}

Substituting in equation (2), yield
\begin{equation}
	E(I)=\alpha \, \text{\^{I}}^{2} \left(\dfrac{1+\gamma\,\text{\^{I}}^{2}}{1+\sigma\,\text{\^{I}}^{2}}\right)
\end{equation}

Therefore, the two-term Bohr-Mottelson formula becomes an extended new formula with three parameters. We denote formula (4) as the collective rotational formula with three parameters (CRF3). The parameters are $ \alpha,\beta,\gamma $.

The suggested CRF3 is more general because it leads to the following three predictions:

a) when  $ \sigma=\gamma $  it gives pure rigid rotor equation(1)

b) when $ \sigma = 0 $ it gives the two parameters Bohr-Mottelson equation (2)

c) when $ \gamma = 0 $ it gives soft rotor model \cite{PhysRevC}

\begin{equation}
	E(I)=\dfrac{\hbar^{2}}{2J} \dfrac{I(I+1)}{1+\sigma (I+I^{2})}
\end{equation}

Two types of moments of inertia were suggested by Bohr-Mottelson which reflect two different aspects of nuclear dynamics. The first moment of inertia is the kinematic $ J^{(1)} $, it is equal to the inverse of the slope of the curve of energy E versus $ \text{\^{I}}^{2} $ (or I(I+1)) times $ \hbar^{2}/2 $, while the second moment of inertia is the dynamic $J^{(2)} $, it is related to the curvature in the curve of E versus \text{\^{I}} (or $ \sqrt{I(I+1) }$ ).

The kinematic $J^{(1)} $) and dynamic $J^{(2)} $ moments of inertia are defined as:

\begin{align}
J^{(1)}&=\dfrac{\hbar^{2}}{2} \left[\dfrac{dE}{dI(I+1)}\right]^{-1}  =\hbar\dfrac{\sqrt{I(I+1)}}{\omega} \nonumber \\
&=\dfrac{\hbar^{2}}{2} \left(\dfrac{dE}{d\text{\^{I}}^{2}}\right)^{-1}=\hbar\dfrac{\text{\^{I}}}{\omega} 
\end{align}
\begin{align}
J^{(2)}&=\hbar^{2}\left[\dfrac{d^{2}E}{d(\sqrt{I(I+1)})^{2}}\right]^{-1}  =\hbar\dfrac{d\sqrt{I(I+1)}}{d\omega} \nonumber \\
&=\hbar^{2} \left(\dfrac{d^{2}E}{d\text{\^{I}} ^{2}}\right)^{-1} =\hbar\dfrac{d\text{\^{I}}}{d\omega} 
\end{align}

In the case of our CRF3, the two moments of inertia becomes
\begin{equation}
	J^{(1)}(I)=\dfrac{\hbar^{2}}{2\alpha} \dfrac{(1+\sigma \text{\^{I}}^{2})^{2}}{[1+\gamma\text{\^{I}}^{2}(2+\sigma\text{\^{I}}^{2})]}
\end{equation}
\begin{equation}
	J^{(2)}(I)=\dfrac{\hbar^{2}}{2\alpha} \dfrac{(1+\sigma \text{\^{I}}^{2})^{3}}{[(1+6\gamma\text{\^{I}}^{2})+\sigma\text{\^{I}}^{2}(3\gamma\text{\^{I}}^{2}+\alpha\gamma\text{\^{I}}^{4}-3)]}
\end{equation}

Experimentally $ \hbar\omega $, $J^{(1)} $and $J^{(2)} $ are extracted in terms of the transition energy $ E_{\gamma}(I)= E(I)-E(I-2) $ as:
\begin{equation}
	\hbar\omega(I) = \frac{1}{4} [E_{\gamma}(I+2) +E_{\gamma}(I)] \;\;\;\;\;\;\;\;\;\;\;\;\;\;\;\;\;(MeV)
\end{equation}
\begin{equation}
	J^{(1)}(I)=\dfrac{2I-1}{E_{\gamma}(I)}      \;\;\;\;\;\;\;\;\;\;\;\;\;\;\;\;\;\;\;\;\;\;\;\;\;\;\;\;\;\;\;\;\;\;\;(\hbar^{2}MeV^{-1})
\end{equation}
\begin{equation}
	J^{(2)}(I)=\dfrac{4}{E_{\gamma}(I+2)-E_{\gamma}(I)}    \;\;\;\;\;\;\;\;\;\;\;\;\;\;\;\;\;  (\hbar^{2}MeV^{-1})
\end{equation}

As a special case, the lowest dynamical moment of inertia reads
\begin{equation}
	J^{(2)}_{lowest} = \dfrac{4}{E_{\gamma}(4^{+}_{1}\rightarrow 2_{1}^{+})-E_{\gamma}(2^{+}_{1}\rightarrow 0_{1}^{+})}
\end{equation}

\section{ Determination of Ground State Band Properties of Even-Even Nuclei and the Physical Identical Parameters}

In order to understand the behavior of low lying states of an axially symmetric normally deformed nuclei, it is insightful to examine some physical observables which exist in a pair of IB's, the observables include:

\textbf{1. The P- Factor, Structure Factor (SF), and Saturation Parameter (SP)}

 Casten \cite{658} introduced the P-Factor
\begin{equation}
	P=\dfrac{N_{p}N_{n}}{N_{p}+N_{n}}
\end{equation}

where $ N_{p} $ and $ N_{n} $ are the numbers of valence protons and valence neutrons respectively which are counted as particles or holes from the nearest closed shell
\begin{align}
N_{p} &= min [(Z-50), (82-Z)] \\
N_{n} &= min [(N-82), (126-N)] 
\end{align}

The P- Factor represents the average number of interactions of each valence nucleon with those of the other type. It can be viewed as the ratio of the number of valences p-n residual interactions to the number of valence like-nucleon pairing interactions, or if the p-n and pairing interactions are orbit independent, then P is proportional to the ratio of the integrated p-n interaction strength to the integrated pairing interaction strength. The nuclear collectivity and deformation depend sensitively on the P- Factor.

The structure factor (SF) and the saturation parameter (SP) are given by
\begin{align}
	SF&=N_{p}N_{n}(N_{p}+N_{n})\\
	SP&=\left(1+\dfrac{SF}{SF_{max}}\right)^{-1}
\end{align}

It is found that the lowest dynamical moment of inertia $ J^{(2)}_{lowest} $ is proportional to $ \sqrt{SF} $.

\textbf{2. The Concept of F-Spin}

A nucleus with $ N_{p} $ valence protons and $ N_{n} $ valence neutrons has a total boson number
\begin{equation}
	N_{B}=\dfrac{N_{p}+N_{n}}{2}=N_{\pi}+N_{\nu}
\end{equation}

The $ N_{\pi} $ proton bosons and neutron bosons are assigned F-Spin, $F= \frac{1}{2} $ with  projection $ F_{0} = + \frac{1}{2} $ for proton bosons and  $ F_{0} = - \frac{1}{2} $ for neutron bosons. A given nucleus is characterized by two quantum numbers \cite{cas}:

$ F=\dfrac{N_{\pi}+N_{\nu}}{2} $ and its projection $ F_{0}=\dfrac{N_{\pi}-N_{\nu}}{2} $

Squaring and subtracting, yield 

\begin{equation}
	4(F^{2}-F^{2}_{0}) = 4 N_{\pi} N_{\nu} = N_{p}N_{n}
\end{equation}

That is any pair of conjugate nuclei with the same F-spin and $ F_{0} $ values in any F-spin multiplet have identical $  N_{p}N_{n} $ values.

In  our chosen nuclei, the F-spin multiplet is given by:
(A+4, Z+2), (A+8, Z+4), (A+12, Z+6) and (A+16, Z+8) for Dy, Er, Yb, Hf, and W isotopes.

Any pair of nuclei which show identical excitation energies have nearly equal value of the product of their valence nucleon numbers $ N_{p} $ and $ N_{n} $ \cite{123}. However, the analysis of experimental data shows that the converse is not true. The simple quantity $ N_{p} N_{n} $ helps also in the evolution of nuclear deformation and collectivity in nuclei \cite{1234}. On the other hand, the product $ N_{p} N_{n} $ or the P- Factor plays an important role in studying the orbit dependence, shell gaps, and intruder orbitals.

\textbf{3. Pairing Interaction Energy}

The pairing interaction energy $ \bigtriangleup $ in an even-even nucleus is the average pairing gap ($( \bigtriangleup_{p} + \bigtriangleup_{n})/ 2$ where $ \bigtriangleup_{p} $ and $ \bigtriangleup_{n} $ are respectively the proton and neutron pairing gaps which are determined from the difference in binding energies of the neighboring odd and even nuclei

\begin{align}
\bigtriangleup_{p} &= \frac{1}{4}[B(N,Z-2)-3B(N,Z-1)+3B(N,Z)-B(N,Z+1)]\\
\bigtriangleup_{n} &= \frac{1}{4}[B(N-2,Z)-3B(N-1,Z)+3B(N,Z)-B(N+1,Z)]
\end{align}

The pairing gaps $ \bigtriangleup_{p} $ and $ \bigtriangleup_{n} $ are determined empirically from the relation
\begin{align}
	\bigtriangleup_{p}\simeq \bigtriangleup_{n} = \dfrac{12}{\sqrt{A}} \texttt{\space}\;\;\;\;\;\;\;\;\;\;\; (MeV)
\end{align}

The average pairing  gap of the nucleus is then

\begin{equation}
	\bigtriangleup=\dfrac{\bigtriangleup_{p}+\bigtriangleup_{n}}{2}=\dfrac{12}{\sqrt{A}}\texttt{\space} MeV
\end{equation}

It is observed that \cite{72,658} the even-even nuclei belong to different mass number having identical $ (N_{p}N_{n}/\bigtriangleup) $ values exhibit identical excitation energies and identical energy ratios.

\textbf{4. Quadrupole Transition Probabilities and Deformation
Parameters}

The quadrupole transition probability per unit time for the
transition $ I_{i} \rightarrow I_{f} $ is given by
\begin{equation}
	T(E_{2}) = \dfrac{4 \pi }{75} \left(\dfrac{5}{\hbar}\right) \left(\dfrac{E_{2^{+}_{1}}}{\hbar c} \right)^{5}B(E_{2};I_{i}\rightarrow I_{f})
\end{equation}

where $ B(E_{2} )$ is the reduced transition probability and $ E_{2^{+}_{1}} $ is the energy of the $ 2_{1}^{+} $ state.

Experimentally $ T(E_{2}) $ for transition $ 2_{1}^{+} \rightarrow 0_{1}^{+} $ is obtained by 
\begin{equation}
T(E_{2},2_{1}^{+} \rightarrow 0_{1}^{+}) = \dfrac{ln 2}{(1+\alpha) T_{1/2}}= \dfrac{0.693}{(1+\alpha) T_{1/2}}
\end{equation}

where $ \alpha $ is the total conversion coefficient taken from the tabulated values given by Rose \cite{me43aref} and $ T_{1/2} $ is the lifetime of the rotational level.

The $ B(E_{2},2_{1}^{+} \rightarrow 0_{1}^{+}) $ values carry important information about the collectivity of nuclear rotation and can be extracted from the equations (25,26).

The relation between the intrinsic nuclear quadrupole moment $ Q_{0} $ and $ B(E_{2}) $ is given by
\begin{equation}
	Q_{0}^{2} =\dfrac{16\pi}{e} B(E_{2},2_{1}^{+}\rightarrow 0_{1}^{+})
\end{equation}

Practically the most reliable method of determining the quadrupole deformation parameter $ \beta_{2} $ in framework of geometric collective model (GCM) is to extract $ \beta_{2} $ from $ Q_{0} $ according to the formula 
\begin{equation}
	\beta_{2}(exp) = \dfrac{\sqrt{5\pi}}{3ZR_{0}^{2}}Q_{0}
\end{equation}

assuming a uniformly charged nucleus of spheroidal shape, where the nuclear radius has the value $ R_{0} = 1.2 A ^{1/3}  $(fm) and Z is the nuclear charge number.

The expression (28) for $ \beta _{2}$ is widely used to compare the quadrupole deformation of different nuclei. It is noticed that the $ B(E_{2},2_{1}^{+} \rightarrow 0_{1}^{+}) $ values increase when going from the closed shell at N=82 toward midshell where maximum values are occur, while from midshell toward the shell closure at N= 126 its values are decreases.

In a second way , specially where the $ B(E_{2},2_{1}^{+} \rightarrow 0_{1}^{+}) $ value is not known, we estimate $ \beta $ by using the approximate empirical Grodzins relation \cite{Ef7}:
\begin{equation} 
	E_{2^{+}_{1}}  B(E_{2},2_{1}^{+} \rightarrow 0_{1}^{+}) = 2.5 \times 10^{-3}\texttt{    } \dfrac{Z^{2}}{A}
\end{equation}

where
\begin{align}
B(E_{2},2_{1}^{+} \rightarrow 0_{1}^{+})= \dfrac{1}{16\pi} e^{2}Q^{2}_{0} =\dfrac{9}{80\pi ^{2}} e^{2}Z^{2}R^{4}_{0}\beta^{2} \;\;\;\;\;\;\;(\texttt{in units of }e^{2}b^{2})
\end{align}

We can relate $ \beta $ and $ E_{2^{+}_{1}} $ as:
\begin{equation}
	\beta^{2}_{G}=\dfrac{1224}{E_{2^{+}_{1}} A^{7/3}}
\end{equation}

where $ E_{2^{+}_{1}} $ is in MeV.

Also $ \beta_{2} $ can be determined by using the SU(3) rotational limit of interacting boson model(IBM)\cite{n12}, the square of the deformation parameter $ \beta^{2} $ in a state of angular momentum I is given by \cite{258}:
\begin{equation}
	\langle\beta^{2} \rangle _{I} = \dfrac{\alpha^{2}}{6(2N-1)} [I(I+1)+8N^{2}_{B}+22N_{B}-15]
\end{equation}
where $ N_{B} $ is the total number of valence bosons and $ \alpha $ is
a normalization constant ($ \alpha= 0.101 $ for rare-earth nuclei). The expectation value of $ \beta^{2} $ in the ground state becomes
\begin{equation}
	\langle\beta^{2}\rangle_{0} = \alpha^{2} \dfrac{8N^{2}_{B}+22N_{B}-15}{6(2N-1)}
\end{equation}

which is an almost linearly increasing function of the boson number $ N_{B} $ and has the same value for nuclei having the same number of valence nucleons
\begin{equation}
	N=[N_{p}+N_{n}],N=[(N_{p}-1)+(N_{n}-1)]
\end{equation}

It is evident that $ \beta_{IBM} $ extracted from IBM is much larger than $ \beta_{GCM} $ extracted from GCM because $ \beta_{GCM} $ refer to the deformation of all A nucleons while $ \beta_{IBM} $ describe only 2N valence bosons, the approximate relation between them is given by:
\begin{equation}
	\beta_{GCM}= 1.18\left( \dfrac{2N}{A}\right)  \beta_{IBM}
\end{equation}

The deformation parameter $ \beta $ reflects the equilibrium shape and structure of the nucleus such as the energy ratio $ R_{4/2}=E(4_{1}^{+})/E(2_{1}^{+}) $ and the reduced transition probability $ B(E_{2}, 2_{1}^{+}\rightarrow 0_{1}^{+}) $ which are the best indicators to exhibit the collective properties of the even-even nuclei.

\textbf{5. Energy Ratios and Percentage Difference in Transition Energies}

The energy ratios and the percentage difference in transition energies give the characteristic of the evolution of the collectivity in the even-even nuclei. Only deformed nuclei show rotational levels and particularly the even-even nuclei display a simple structure energies proportional to I(I+1) with only even values of the spin I considering that the moment of inertia is constant (rigid rotator), therefore the energy ratio $ R_{4/2} = 3.333 $. The observed 
moment of inertia extracted from the experiment is only one-quarter to one-half of what one would expect from a rigid rotator which means that not the whole nucleons are participating in the collective motion.
 
On the other hand for an ideal harmonic quadrupole spectrum for spherical nuclei a system of equidistant states is formed by the composition of vibrational quanta. The first excited state is $ 2_{1}^{+} $ followed by the degenerate $ 0_{2}^{+},2_{2}^{+},4_{1}^{+}, $ and  so forth. Therefore energy ratio$  R_{4/2}= 2 $.

To compare level spacing in two nuclei with masses $ A_{1} $, and $ A_{2} $ where $ A_{2} > A_{1} $,  we define the percentage differences ratios in transition energies as :
\begin{equation}
	\delta =\dfrac{\bigtriangleup E_{\gamma}(I)}{E_{\gamma_{2}}(I)}
\end{equation}

where
 \begin{align}
E_{\gamma} = E(I)-E(I-2)\\
\bigtriangleup E_{\gamma}(I)=E_{\gamma_{1}}(I)-E_{\gamma_{2}}(I)
 \end{align}
So that
\begin{align}
E_{\gamma_{1}}=(1+\delta)E_{\gamma_{2}}
\end{align}
For rigid rotor the ratio 
\begin{align}
\delta_{R}= \left(\dfrac{A_{2}}{A_{1}}\right)^{5/3}-1
\end{align}

define the fractional change in $ A^{5/3} $.

The fractional change in transition energies $ \delta $ divided by the rigid rotor  ratio $ \delta_{R} $ is denoted by $ \delta_{\gamma} $. If the spacings are identical, then $ \delta=0,\delta_{\gamma}=0 $ and if they scale as $ A^{5/3} $ then $ \delta_{\gamma=1} $.

Similarly, the percentage difference in kinematic moment of inertia $ J^{(1)} $ is given by

\begin{align}
	K= - \dfrac{\bigtriangleup J^{(1)}(I)}{J^{(1)}_{2}(I)}
\end{align}
where
\begin{align}
	J^{(1)}(I)&= \dfrac{2I-1}{E_{\gamma}(I)}\\
	\bigtriangleup J^{(1)}(I)&=  J^{(1)}_{1}(I) - J^{(1)}_{2}(I)
\end{align}

So that
\begin{equation}
	J^{(2)}_{2} =(1+K)  J^{(1)}_{1}
\end{equation}

Substituting for $  J^{(1)} $, yield $ K=\delta $.

\section{The Interacting Boson Model to Calculate the Potential Energy Surfaces and Electric Quadrupole Transition Probability}

We consider the Hamiltonian of the first order U(5)- SU(3) quantum shape phase transition in the form
\begin{equation}
	H=\epsilon _{d} \hat{n} _{d} +a_{2} \hat{Q}^{(x)}\hat{Q}^{(x)}
\end{equation}

where $ \hat{n} _{d} $ and $ \hat{Q}^{(x)} $ are respectively the d-boson number operator and quadrupole operator defined as
\begin{align}
	\hat{n} _{d} &=\sum_{\mu} d_{\mu}^{\dagger} \stackrel{\sim}{d}_{\mu} \\
	\hat{Q}^{(x)}&=\left[d^{\dagger}s+s^{\dagger}\stackrel{\sim}{d}\right]^{(2)}+x\left[d^{\dagger}\times\stackrel{\sim}{d}\right]^{(2)}
\end{align}

where $ \left(s^{\dagger},d^{\dagger}\right) $ and $ \left(s,\stackrel{\sim}{d}\right) $ are the boson creation and annihilation operators respectively, and $  x $ is the structure parameter of the quadrupole operator of IBM  ($ x $ for pure rotational SU(3) limit is equal to $ -\sqrt{7}/2 $). Here  $ d_{\mu} = (-1)^{\mu} d_{-\mu} $ and standard notation of angular momentum coupling is used.

To get the potential energy surface (PES) of the Hamiltonian, we introduce the intrinsic coherent frame in which the ground state of a nucleus with N bosons can be expressed as a boson condensate. The bosonic intrinsic coherent state for the ground  state band of a given even-even nucleus can be written in the form\cite{2,1,3}

\begin{equation}
	\lvert N\beta\gamma\rangle =\dfrac{1}{\sqrt{N!}} [b^{\dagger}(\beta,\gamma)]^{N} \lvert 0\rangle
\end{equation}
where $ \lvert0\rangle $ is the boson vacuum and $ b^{\dagger} $ is the boson creation operator which acts in the intrinsic system and is given by:
\begin{align}
	b^{\dagger} =\dfrac{1}{\sqrt{1+\beta^{2}}} [s^{\dagger}+\beta cos\gamma(d_{0}^{\dagger})+\dfrac{1}{\sqrt{2}}\beta sin\gamma(d_{2}^{\dagger}+d_{-2}^{\dagger})]
\end{align}

where $ \beta $ is the quadrupole deformation parameter which measures the axial deviation from spherical symmetry and the parameter $ \gamma $ controls the departure from axial symmetries.

The ground state PES is the expectation value of the Hamiltonian in the intrinsic coherent state

\begin{equation}
	PES=\langle N\beta\gamma\rvert H\rvert N\beta\gamma\rangle
\end{equation}
The associated PES of the Hamiltonian (45) for $ x=-\sqrt{7}/2 $ reads

\begin{align}
	E(N,\beta,\gamma) &=\epsilon_{d} \dfrac{N\beta^{2}}{1+\beta^{2}}+a_{2}\left[\dfrac{N}{1+\beta^{2}}(5+\dfrac{11}{4}\beta^{2})+\dfrac{N(N-1)}{(1+\beta^{2})^{2}}(4\beta^{2}-2\sqrt{2}\beta^{3}cos3\gamma+\dfrac{1}{2}\beta^{4})\right]
\end{align}

Equation (51) can be written in another form as
\begin{align}
	E(N,\beta,\gamma) = g_{1}\dfrac{N\beta^{2}}{1+\beta^{2}}+\dfrac{N(N-1)}{(1+\beta^{2})^{2}}[g_{2}\beta^{2}+g_{3}\beta^{3}cos3\gamma+g_{4}\beta^{4}]+c
\end{align}
where the coefficients involve linear combination of the Hamiltonian parameters

\begin{align*}
	g_{1}&=\epsilon_{d}-\dfrac{9}{4}a_{2} ,\;\;\;\;\;\;\;\;\;\; g_{2}=4a_{2}\\
	g_{3}&=2\sqrt{2}a_{2},\;\;\;\;\;\;\;\;\;\;\;\;\;\; g_{4}=\dfrac{1}{2}a_{2},\;\;\;\;\;\;\;c=5Na_{2}
\end{align*}

Also, equation (51) can be rewritten in general form as
\begin{equation}
	E(N,\beta,\gamma) = \dfrac{A_{2}\beta^{2}+A_{3}\beta^{3}cos3\gamma+A_{4}\beta^{4}}{(1+\beta^{2})^{2}}+A_{0}
\end{equation}
where the coefficients read
\begin{align*}
	A_{2}&=\left[\epsilon+\left(4N-\dfrac{25}{4}\right)a_{2}\right]N ,\;\;\;\;\;\;\;\;\;\;\;A_{3}=2\sqrt{2}a_{2}(N-1)N \\
	A_{4}&=\left[\epsilon+\left(\dfrac{2N+5}{4}-4\right) a_{2}\right]N,\;\;\;\;\;\;A_{0}=5a_{2}N
\end{align*}

For $ a_{2}=0 $, we get the pure spherical vibrator U(5) limit and for $ \epsilon_{d} = 0 $, we get the pure deformed rotational Su(3) limit.

Another important quantity that tests the nature of the shape phase transition of low lying collective states  the reduced electric quadrupole transition probabilities $ B(E_{2}) $.

In IBM, the general form of the electric quadrupole operator is written in the form \cite{155}
\begin{equation}
	T(E_{2})=e Q(sdIBM)
\end{equation}

The coefficient e is the boson's effective charge.

The reduced electric quadrupole transition probabilities are given by

\begin{equation}
	B[E_{2},I_{i}\rightarrow I_{f}] = \dfrac{1}{2I_{i}+1}\rvert\langle I_{f}\rvert\rvert T(E_{2})\rvert\rvert I_{i}\rangle\rvert^{2}
\end{equation}

For rotational SU(3), yield

\begin{align}
	B(E_{2},I+2\rightarrow I)&=e^{2}\,\dfrac{3}{4}\dfrac{(I+2)(I+1)}{(2I+3)(2I+5)}(2N-1)(2N+I+3)\\
	Q(I)&=-e\sqrt{\dfrac{16\pi}{40}} \dfrac{I}{2I+3}(4N+3)
\end{align}

For the special case for I=0, we have
\begin{align}
	B(E_{2}, 2_{1}^{+} \rightarrow 0_{1}^{+}) = e^{2} \dfrac{1}{5}N(2N+3)
\end{align}

\section{Numerical Calculations and Discussion}

In this section, we applied our formalism to eight pairs of nuclei having identical bands (IB's) in rare-earth region namely:  $ ( ^{162}Yb- ^{166}Hf), ( ^{162}Er- ^{166}Yb), ( ^{162}Dy- ^{166}Er),(^{160}Dy - ^{168}Yb), (^{160}Er- ^{168}Hf), \\(^{158}Er- ^{170}W), (^{158}Dy- ^{170}Hf) $ and $ ( ^{156}Dy- ^{172}W) $.

To calculate the ground state positive parity excitation energy E(I) for each nucleus, we suggested the CRF3.

The parameters $ \alpha,\gamma, \sigma$ of CRF3 have been determined by a fitting procedure using a computer-simulated search program to minimize the root mean square deviation of the calculated excitation energies from the experimental ones. The quality of the fitting is indicated by the standard common definition of $ x $
\begin{align*}
	x=\sqrt{\dfrac{1}{N}\Sigma_{i}\left(\dfrac{E_{exp}(I_{i})-E_{cal}(I_{i})}{\delta E_{exp}(I_{i})}\right)^{2}}
\end{align*}

where N is the number of experimental data points entering the fitting procedure and  $ \delta E_{exp}(I_{i}) $ is the experimental error in the excitation energies - The experimental excitation energies are taken from \cite{050}. The optimized best adopted values of parameters for each nucleus of our studied nuclei are listed in Table (\ref{tab:1}).
\begin{figure}[H]
	\centering
	\includegraphics[height=.8\textheight]{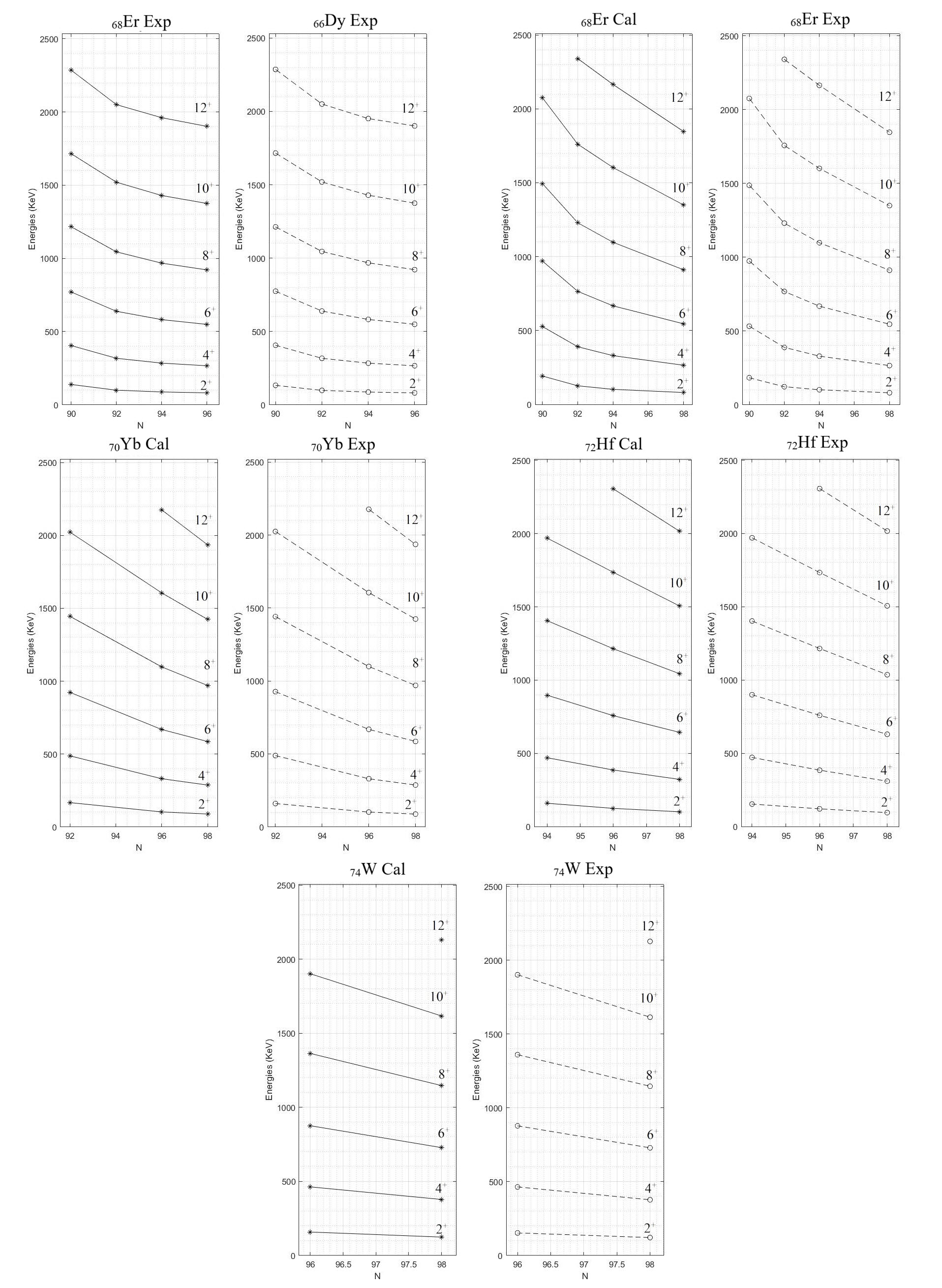}
	\caption{Systematic of the calculated (solid curves) ground state energies for our selected even-even rare earth Dy, Er, YB, Hf, W  isotopes versus neutron number N and comparison with the experimental ones (dashed curves). The spin-parity are labeled by $I^{\pi}$.  }
	\label{f1}
\end{figure}

\begin{longtable}[H]{c||c|c|c|c|c}
	\caption{Values of optimized best parameters $ \alpha,\gamma,\sigma $ of the collective rotational formula(CRF3) for ground state bands in our selected even-even rare-earth nuclei. $ N_{p} $ and $ N_{n} $ are the number of valance protons and the number of valance neutrons respectively.}
	\label{tab:1}  \\     
	
		\hline\noalign{\smallskip}
		$ \;\;\;\;\; $Nuclide$ \;\;\;\;\; $ &$ \;\;\;\;\; $ $ \alpha $ (KeV) $ \;\;\;\;\; $&$ \;\;\;\;\; $ $ \gamma $ ($ 10^{-3} $)  $ \;\;\;\;\; $&$ \;\;\;\;\; $ $ \sigma $ ($ 10^{-3} $) $ \;\;\;\;\; $&$ \;\;\;\;\; $ $ N_{p} $ $ \;\;\;\;\; $&$ \;\;\;\;\; $ $ N_{n} $ $ \;\;\;\;\; $ \\
		\noalign{\smallskip}\hline\noalign{\smallskip}
		Dy 156 & 22.96 &6.964 &14.54 &16 & 8 \\
		$ \;\;\;\;\;\; $158 & 16.48 &2.163 &4.339 &16 & 10 \\
		$ \;\;\;\;\;\; $160 & 14.49 &0.8683 &2.021 &16 & 12 \\
		$ \;\;\;\;\;\; $162 & 13.49 &1.398 &2.233 &16 & 14 \\ 
		&&&&&\\
		Er 158 & 32.76 &9.699 &23.52 &14 & 8 \\
		$ \;\;\;\;\;\; $160 & 20.73 &3.017 &6.641 &14 & 10 \\
		$ \;\;\;\;\;\; $162 & 17.01 &1.440 &3.212 &14 & 12 \\
		$ \;\;\;\;\;\; $166 & 13.49 &0.2573 &1.188 &14 & 16 \\
		&&&&&\\
		Yb 162 & 27.87 &6.334 &14.27 &12 & 10 \\
		$ \;\;\;\;\;\; $166 & 17.08 &2.053 &3.95 &12 & 14 \\
		$ \;\;\;\;\;\; $168 & 14.72 &1.039 &2.425 &12 & 16 \\
		&&&&&\\
		Hf 166 & 26.60 &5.565 &12.67 &10 & 12 \\
		$ \;\;\;\;\;\; $168 & 20.58 &3.116 &6.849 &10 & 14 \\
		$ \;\;\;\;\;\; $170 & 15.92 &-0.00749 &1.391 &10 & 16 \\
		&&&&&\\	
		W $ \; $170 & 26.44 &5.714 &13.55 &8 & 14 \\
		$ \;\;\;\;\;\; $172 & 20.68 &3.944 &9.279 &8 & 16 \\
		
		\noalign{\smallskip}\hline
	
\end{longtable}

\begin{figure}[H]
	\centering
	\includegraphics[height= .32\textheight]{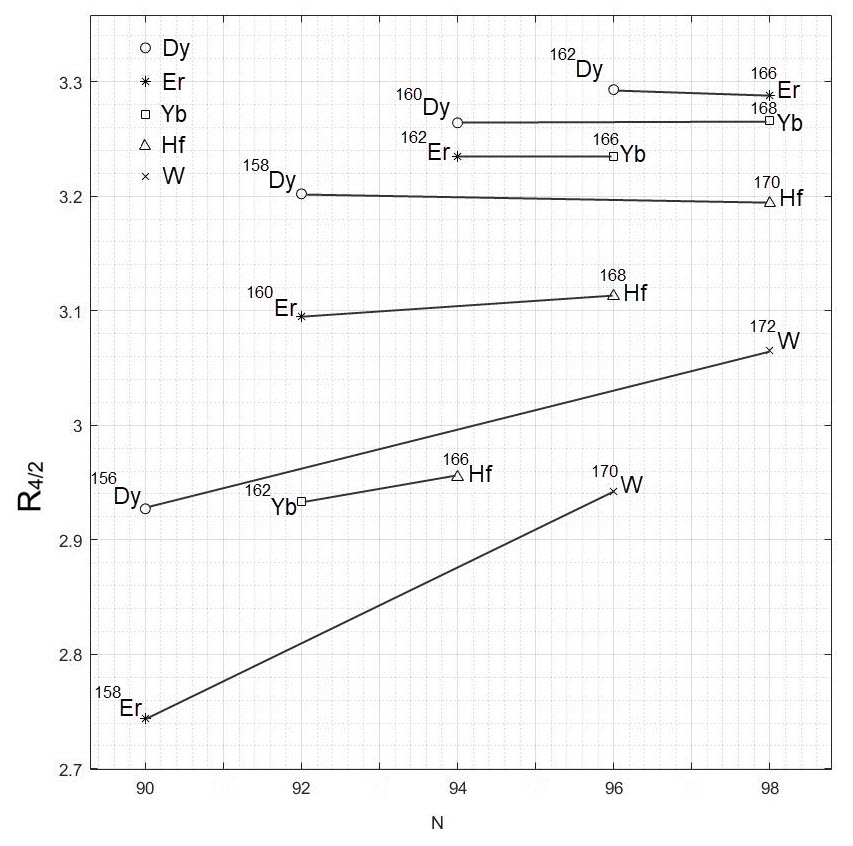}
	\caption{The calculated energy ratio $R_{4/2}=E(4^{+}_{1})/E(2^{+}_{1})$ versus neutron number N characterizes the low lying spectrum in Dy, Er, Yb, Hf, and W isotopes. The symbols $o, \ast, \Square, \triangle,$ and x denote $_{66}Dy, _{68}Er, _{70}Yb, _{72}Hf, $ and $ _{74}W $ respectively.}
	\label{f2}
\end{figure}

The systematic of the excitation energies of the low spin states as a function of neutron number N in the considered even-even Dy, Er, Yb, Hf, W isotopes in the mass region A= 156 - 172 in the normally deformed nuclear are shown in Figure(\ref{f1}) and compared with the experimental ones. Only the ground state of positive parity and spin $I^{\pi}=2^{+},4^{+},6^{+},8^{+},10^{+},12^{+} $ has been indicated. We can see that the excitation energies decrease with increasing the neutron number. Also, Figure(\ref{f2}) illustrate the calculated energy ratio $ R_{4/2} $ as a function of neutron number N for our studied nuclei. We observe that for each isotopic chain the value of $ R_{4/2} $ increases with increasing N (that is the deformation increased), and the difference in $ R_{4/2} $ for all pairs of IB's is ranging from 0.4 \% to 2.5 \% except the two pairs including the two isotopes $ ^{170,172}W $ (the difference is about 5\%).

\begin{figure}[H]
	\centering
	\includegraphics[width= \textwidth]{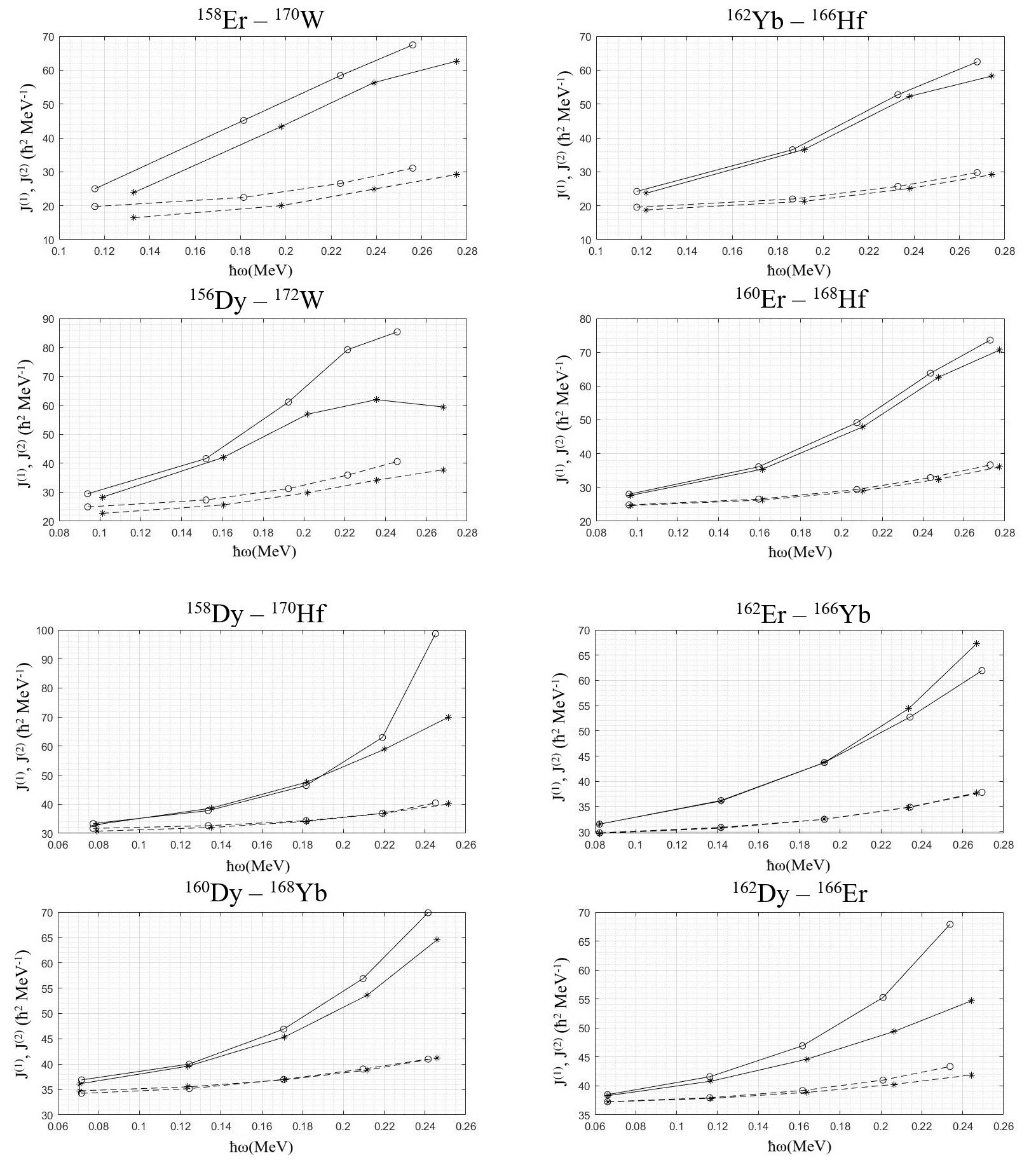}
	\caption{The calculated results of kinematic $J^{(1)}$ (dashed curves) and dynamic $J^{(2)}$ (solid curves) moments of inertia plotted as a function of rotational frequency  $\hbar\omega$  for the studied eight pairs of identical bands in the rare-earth region. The $\ast$ and o correspond to the lighter and heavier nucleus respectively.}
	\label{f3}
\end{figure}

For the eight pairs of IB'S, the kinematic $ J^{(1)} $ and the dynamic $ J^{(2)} $  moments of inertia derived from the transition energies are plotted versus the rotational frequency $ \hbar \omega$ as shown in Figure(\ref{f3}). It can be seen that for all bands $ J^{(1)} $ is smaller than $ J^{(2)} $ and a smooth gradual increase in both $ J^{(1)} $ and $ J^{(2)} $  with increasing $ \hbar\omega $ are seen and the similarities between each pair of IB'S are observed.

The IB's correlation quantities exist between the considered pairs of nuclei which exhibit the same identical excitation energies in their ground state bands are listed in Table (\ref{tab:21}). These quantities include the P.  Factor, structure Factor SF, Saturation parameter SP, the F-Spin and its projection $ F_{0}$, pairing gaps $ \bigtriangleup $, and the deformation parameter $ \beta $. The maximum structure factor for our region of nuclei is SF= 6720. It is seen that the ratio $ N_{p}N_{n}/\bigtriangleup $ rather than the product $ N_{p}N_{n} $ may be a better parameter for studying the IB's. Note that nuclei with symmetric $ \pm F_{0} $ values have identical $ N_{p}N_{n }$ values. For example the pair ($  ^{160}Er$ and $^{168}Hf $) have $ (N_{p},N_{n})=(14,10) $ and $ (10,14) $ respectively, so that $ N_{p}N_{n}=140 $  and $ F_{0}=\pm 1 $. Therefore if any F-spin multiplet has $ F_{0}=\rvert N_{p}-N_{n}\rvert/4 $, those indicate that the pair of nuclei are similar in structure if they have identical $ (\rvert F_{0}\rvert,N_{p}N_{n}) $.

\begin{longtable}[h!]{c||c|c|c|c|c|c}
	\caption{The identical band quantities of our eight pairs of nuclei.}	\label{tab:21}\\
	\hline\noalign{\smallskip}
	 & $ N_{p}N_{n} $ & P & SF & SP & $ \lvert\delta\rvert \% $ & $ \lvert k\rvert \%  $\\
	
	\hline\noalign{\smallskip}
	
	
	
	($ ^{158}Er  \;-\;  ^{170}W \; $) & 112 & 5.090 & 2464 & 0.7317 & 1.28 & 1.27\\
	($ ^{162}Yb  \;-\;  ^{166}Hf  $) & 120 & 5.4545 & 2640 & 0.7179 & 2.94 & 2.45\\
	($ ^{156}Dy  \;-\;  ^{172}W  \;$) & 128 & 5.333 & 3072 & 0.6862 & 6.73 & 6.28\\
	($ ^{160}Er  \;-\;  ^{168}Hf  $) & 140 & 5.833 & 3360 & 0.6666 & 1.35 & 1.22\\
	($ ^{158}Dy  \;-\;  ^{170}Hf  $) & 160 & 6.1538 & 4160 & 0.6176 & 1.28 & 1.27\\
	($ ^{162}Er  \;-\;  ^{166}Yb  $) & 168 & 6.6461 & 4368 & 0.6060 & 0.22 & 0.20\\
	($ ^{160}Dy  \;-\;  ^{168}Yb  $)  & 192 & 6.6857 & 5376 & 0.5555 & 0.10 & 0.30\\
	($ ^{162}Dy  \;-\;  ^{166}Er  $) & 224 & 7.466 & 6720 & 0.5 & 1.29 & 1.26\\
	\hline\noalign{\smallskip}
\end{longtable}

\begin{longtable*}[h!]{c||c|c|c|c|c|c|c}
	\hline\noalign{\smallskip}
	& $ (N_{\pi},N_{\nu}) $ & N & $ \dfrac{N_{\nu}}{N_{\pi}} $ & $ (F,F_{0}) $ & $ \bigtriangleup $ (MeV) & $ \dfrac{N_{p}N_{n}}{\bigtriangleup}  $(MeV$ ^{-1} $)& $ \beta_{G} $\\
	
	\hline\noalign{\smallskip}
		\hline\noalign{\smallskip}
	
	

	 $\;\; ^{158}Er \;\;\;\; $ & (7,4) & 11 & 0.571 & (5.5,1.5) & 0.954 & 117.4 & 0.2173 \\
	 $\;\; ^{170}W   \;\;\;\;$ & (4,7) & 11 & 1.750 & (5.5,-1.5) & 0.920 & 121.739 & 0.2206 \\
	\hline\noalign{\smallskip}
	 $\;\; ^{162}Yb  \;\;\;\;$ & (6,5) & 11 & 0.833 & (5.5,0.5) & 0.942 & 127.388 & 0.2270 \\
	 $\;\; ^{166}Hf    \;\;\;\;$ & (5,6) & 11 & 1.2 & (5.5,-0.5) & 0.931 & 128.893 & 0.2254 \\
	\hline\noalign{\smallskip}
	 $\;\; ^{156}Dy    \;\;\;\;$ & (8,4) & 12 & 0.5 & (6,2) & 0.960 & 133.333 & 0.2601 \\
	 $\;\;\;\; ^{172}W     \;\;\;\;$ & (4,8) & 12 & 2.0 & (6,-2) & 0.914 & 140.043 & 0.2459 \\
	\hline\noalign{\smallskip}
	 $\;\; ^{160}Er  \;\;\;\;$ & (7,5) & 12 & 0.714 & (6,1) & 0.948 & 147.679 & 0.2643 \\
	 $\;\; ^{168}Hf    \;\;\;\;$ & (5,7) & 12 & 1.4 & (6,-1) & 0.925 & 151.351 & 0.2517 \\
	\hline\noalign{\smallskip}
	 $\;\; ^{158}Dy  \;\;\;\;$ & (8,5) & 13 & 0.625 & (6.5,1.5) & 0.954 & 167.714 & 0.3026 \\
	 $\;\; ^{170}Hf   \;\;\;\; $ & (5,8) & 13 & 1.6 & (6.5,-1.5) & 0.920 & 173.913 & 0.2754 \\
	\hline\noalign{\smallskip}
	 $\;\; ^{162}Er  \;\;\;\;$ & (7,6) & 13 & 0.857 & (6.5,0.5) & 0.942 & 178.343 & 0.2896 \\
	 $\;\; ^{166}Yb  \;\;\;\;$ & (6,7) & 13 & 1.166 & (6.5,-0.5) & 0.931 & 180.451 & 0.2814 \\
	\hline\noalign{\smallskip}
	 $\;\; ^{160}Dy  \;\;\;\; $ & (8,6) & 14 & 0.75 & (7,1) & 0.948 & 202.531 & 0.3181 \\
	 $\;\; ^{168}Yb \;\;\;\; $ & (6,8) & 14 & 1.333 & (7,-1) & 0.925 & 207.567 & 0.2993 \\
	\hline\noalign{\smallskip}
	 $\;\; ^{162}Dy \;\;\;\; $ & (8,7) & 15 & 0.875 & (7.5,0.5) & 0.942 & 237.791 & 0.3256 \\
	 $\;\; ^{166}Er \;\;\;\;$ & (7,8) & 15 & 1.142 & (7.5,-0.5) & 0.931 & 240.601 & 0.3167 \\
	 
	\hline\noalign{\smallskip}
\end{longtable*}

The percentage differences ratios in transition energy $ \delta $ and the rigid rotor ratio $ \delta_{R} $  between pairs of levels in two nuclei are calculated and listed in Table(\ref{tab:3}) for our eight pairs of IB's. In spite of the parameters $ N_{p}N_{n} $, P, SF and SP are the same for the pairs $ (^{156}Dy, ^{172}W) $, this pair is not really identical according to their high average percentage differences in transition energies (approximately 6.7\%).


	
	
	
	
	
\begin{figure}[H]
	\centering
	\includegraphics[height= .4\textheight]{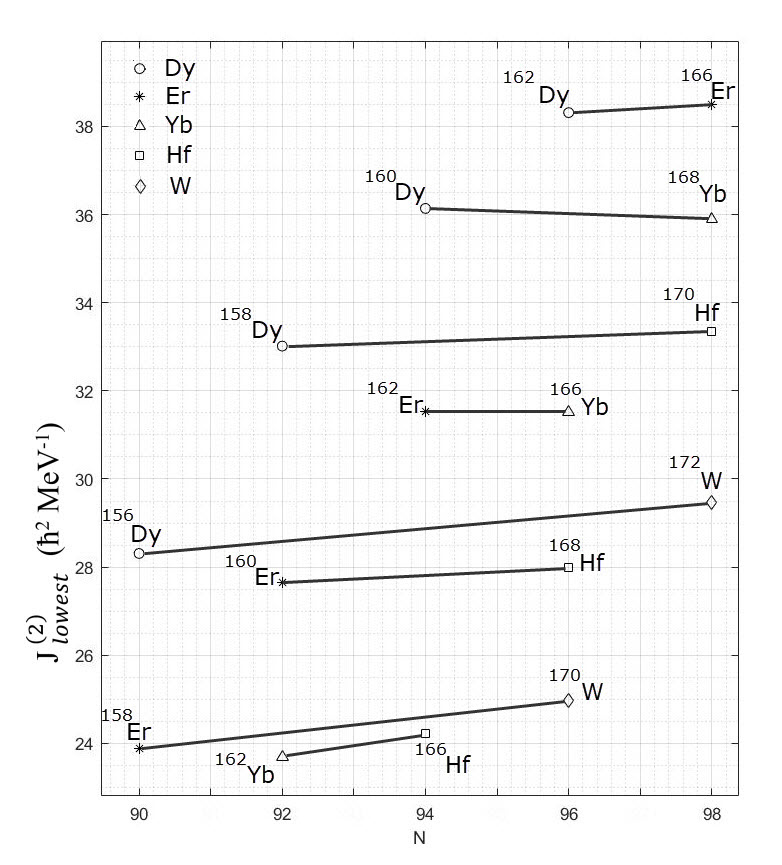}
	\caption{The lowest dynamical moment of inertia $J^{(2)}_{lowest}$ against the neutron number N for the eight pairs of identical bands. The solid line connects each pair and symbols $o, \ast, \triangle, \Square,$ and $\diamondsuit $ denotes  $_{66}Dy, _{68}Er, _{70}Yb, _{72}Hf, $ and $ _{74}W $ respectively.
	}
	\label{f4}
\end{figure}

For each nucleus in isotopic chains of $ _{66}Dy,_{68}Er,_{70}Yb,_{72}Hf $ and $ _{74}W $, the values of lowest dynamical moments of inertia $ J^{(2)}_{lowest} $ were calculated and displayed against the neutron number N in Figure(\ref{f4}) - It can be seen that $ J^{(2)}_{lowest} $ increases with increasing the neutron number N and the difference in$ J^{(2)}_{lowest} $  for each pair of IB's is very small ( approximately a horizontal line). As an example of two nuclei that exhibit good IB's, the pair $ ^{162}_{68}Er( J^{(2)}_{lowest}=31.525 \hbar^{2}MeV^{-1} ) $ and $ ^{166}_{70}Yb ( J^{(2)}_{lowest}=31.519 \hbar^{2}MeV^{-1} ) $, that is nearly the same $ J^{(2)}_{lowest} $.

\begin{table}[H]
	\caption{The percentage differences ratios in transition energies $ \delta $, the fractional change in transition energies divided by the rigid rotor ratio $ \delta R $ and the ratio $ R \; = \delta/\delta R  $ for the eight pairs of identical bands.}	
	\label{tab:3}
	\centering
	\begin{tabular}{c||c|c|c}
		\hline\noalign{\smallskip}
		Identical pairs & $ \lvert \delta\rvert =\dfrac{\bigtriangleup E_{\gamma}}{E_{\gamma_{2}}} \;\; \% $ & $ \delta R $ &  $ \langle R_{\delta}\rangle $\\
		\noalign{\smallskip}\hline\noalign{\smallskip}
		($ ^{162}Yb  \;-\;  ^{166}Hf  $) & 2.964 & 4.149 & 0.714 \\
		($ ^{162}Er  \;-\;  ^{166}Yb  $) & 0.415 & 4.149 & 0.100 \\
		($ ^{162}Dy  \;-\;  ^{166}Er  $) & 1.297 & 4.149 & 0.312 \\
		($ ^{160}Er  \;-\;  ^{168}Hf  $) & 1.352 & 8.471 & 0.159 \\
		($ ^{160}Dy  \;-\;  ^{168}Yb  $) & 1.131 & 8.471 & 0.133 \\
		($ ^{158}Er  \;-\;  ^{170}W  \;$) & 10.826 & 12.976 & 0.834 \\
		($ ^{158}Dy  \;-\;  ^{170}Hf  $) & 1.765 & 12.976 & 0.136 \\
		($ ^{156}Dy  \;-\;  ^{172}W \; $) & 7.410 & 17.671 & 0.419 \\
		\noalign{\smallskip}\hline
	\end{tabular}
\end{table}

\begin{figure}[H]
	\centering
	\includegraphics[width=.9\textwidth]{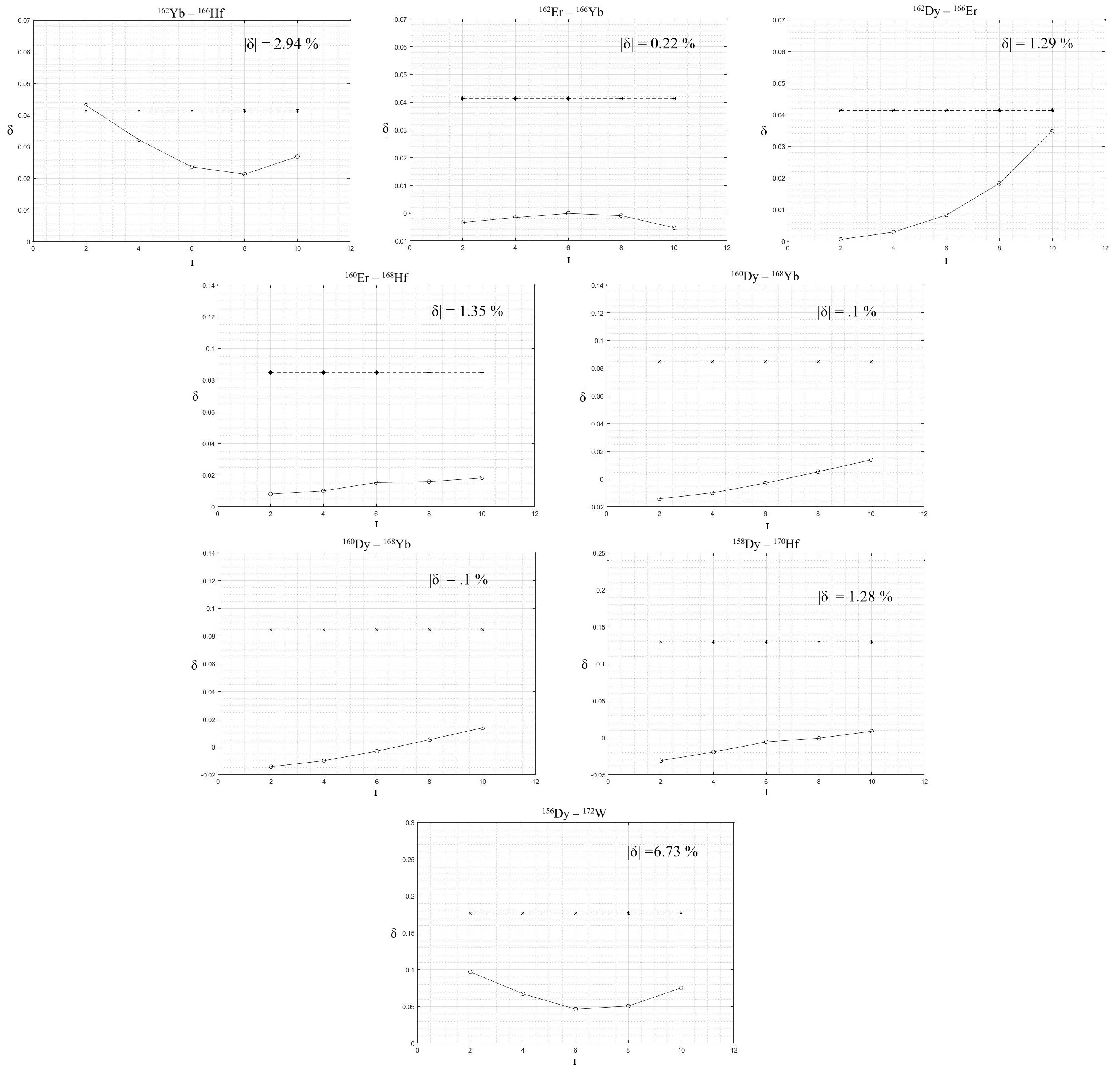}
	\caption{Percentage difference in transition energies $ \delta=\bigtriangleup E_{\gamma}/E_{\gamma_{2}} $ for the eight pairs of multiplet (A+4,Z+2), (A+8,Z+4), (A+12,Z+6), and (A+16,Z+8) for Dy, Er, Yb, Hf, and W isotopes. The dashed curve represents the ratio of the rigid rotor.}
	\label{f5}
\end{figure}

We classified our selected pairs of IB's into four multiplets = (A+4),  Z+2), (A+B,Z+4), (A+12,Z+6), and (A+16,Z+8) and the percentage differences in transition energies $ \delta=\bigtriangleup E_{\gamma}/E_{\gamma_{2}} $ as a function of spin I (up to I=10) have been calculated and illustrated Figure (\ref{f5}). It is seen that the pairs of IB's have approximately similar $ \delta $ ( less than 2.5 \%) except the two pairs which include the tungsten isotopes $ ^{170,172}W $ where the value of $ \delta $ reaches $  \sim 6-10 $\% in spite of they have the same $ N_{p}N_{n} $ value ($ N_{p}N_{n}=112 $ for $ ^{158}Er, ^{170}W $ and $ N_{p}N_{n}  = 128 $ for $ ^{156}Dy,^{172}W $).

\begin{figure}[H]
	\centering
	\includegraphics[width=.5\textwidth]{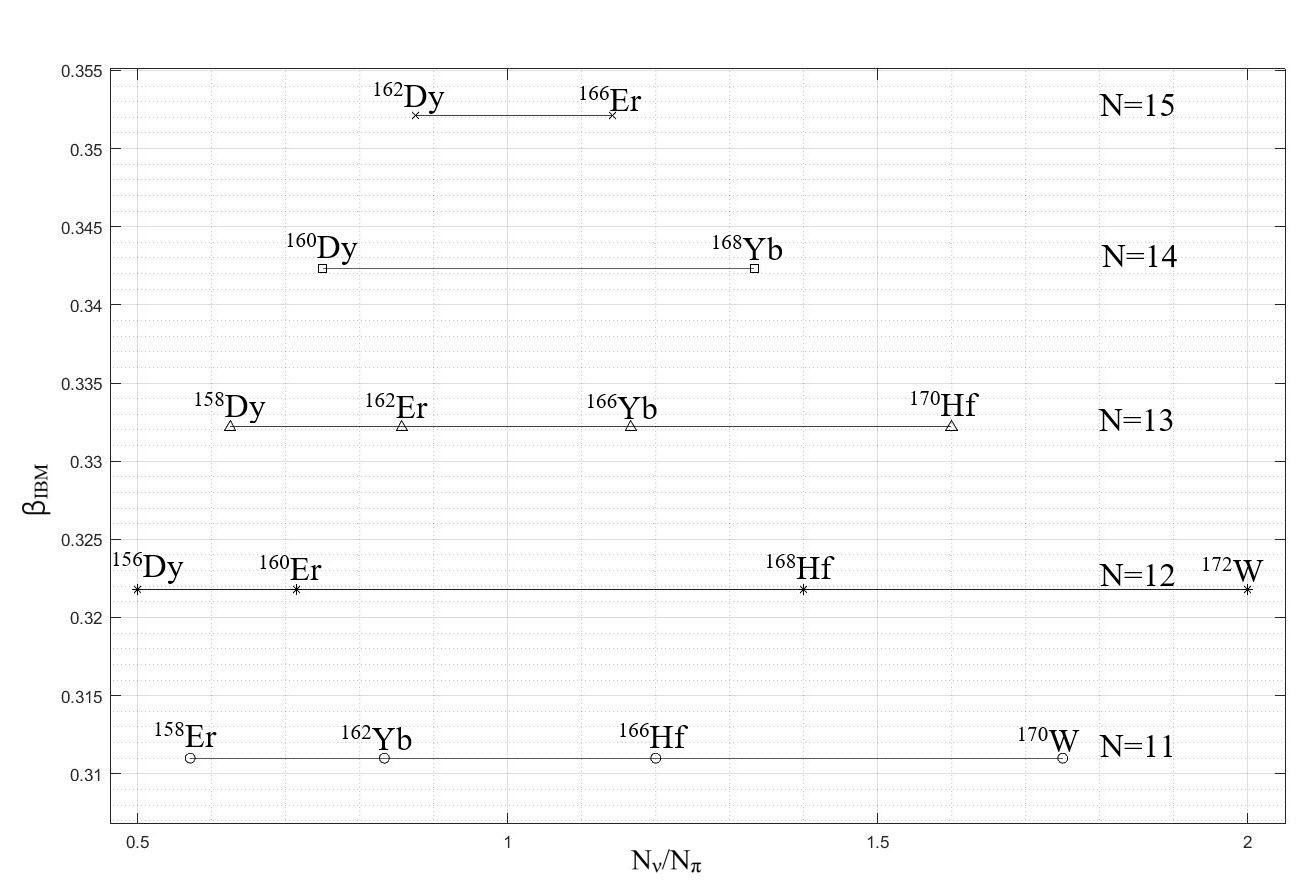}
	\caption{The quadrupole deformation parameter $ \beta_{IBM} $ was calculated from SU(3) limit of IBM as a function of $ N_{\nu}/N_{\pi} $ for our eight pairs of identical bands.}
	\label{f6}
\end{figure}

To further investigation for IB's we used the SU(3) rotational limit of the IBM to extract the quadrupole deformation $ \beta_{IBM} $ for each nucleus. The calculated $ \beta_{IBM} $ is plotted against the ratio $ N_{\nu}/N_{\pi} $  (where $ N_{\nu} $ and $  N_{\pi} $ are the number of valence neutron and valence proton bosons respectively) in Figure(\ref{f6}). It is seen that $ \beta_{IBM} $ is the same for each pair of IB's (horizontal line).


\begin{figure}[H]
	\centering
	\includegraphics[height=.78\textheight]{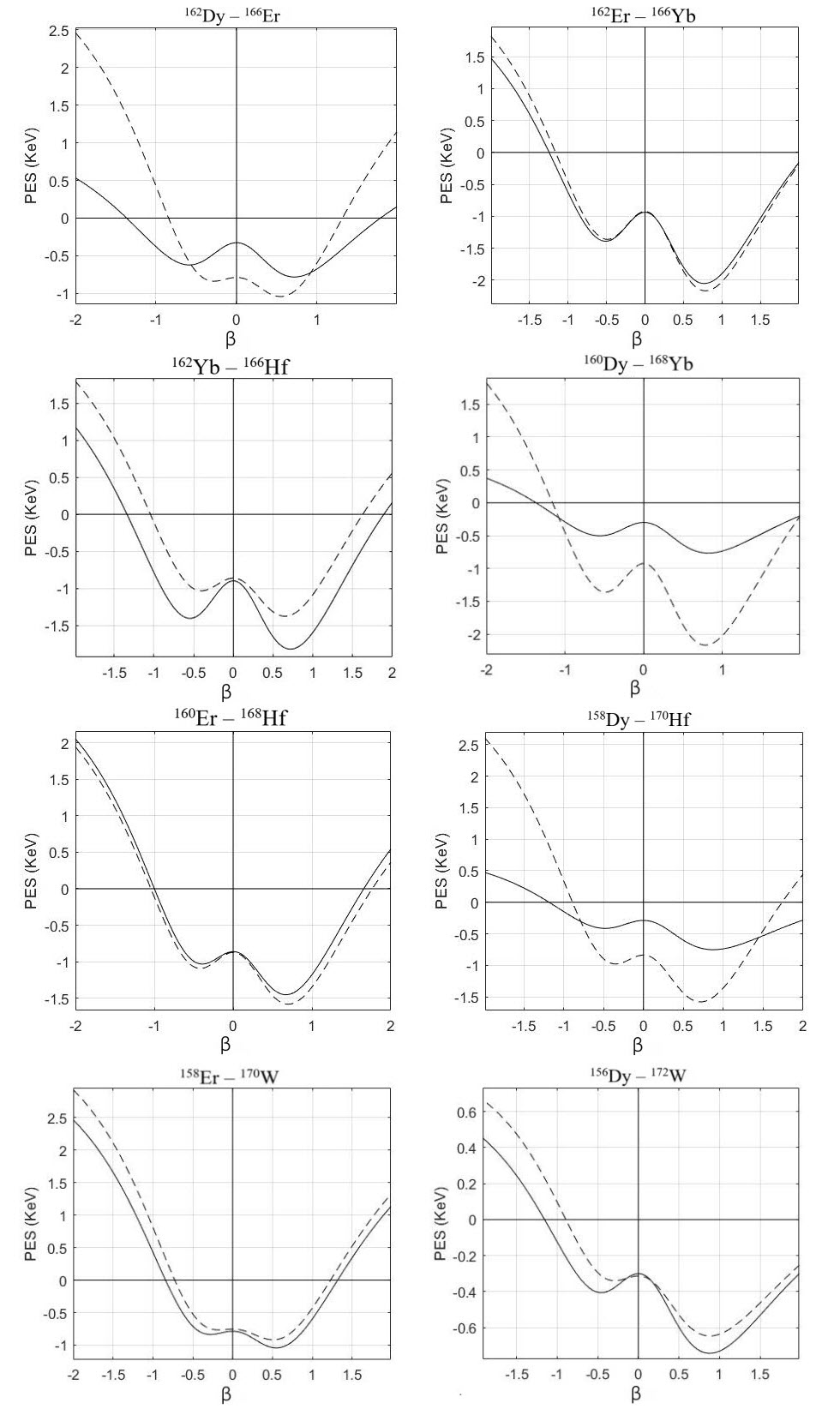}
	\caption{Sketch of the potential energy surface PES calculated from the U(5)-SU(3) shape phase transitions of IBM with intrinsic coherent state versus the deformation parameters $\beta$ for the eight pairs of even-even nuclei having identical bands. }
	\label{f7}
\end{figure}

For each nucleus, by using the IBM Hamiltonian equation (45) and its eigenvalues equation (53), the PES's have been calculated as a function of deformation parameter $ \beta $ along the axial trajectory $ \gamma$ = 0\textdegree, 60\textdegree . The results are illustrated in Figure(\ref{f7}) and the corresponding calculated parameter of the PES's $ A_{2}, A_{3}, A_{4} $ and $ A_{o} $ which are linear combinations of the original parameters $ \epsilon_{0} $ and $ a_{2} $ are listed in Table(4). From the graphs presented in Figure(\ref{f7}), we observe the similarity in PES's for each pair of IB's. All studied nuclei are deformed and have rotational characters, the prolate deformation is deeper than the oblate deformation.

\begin{longtable}[T]{c||c|c|c|c|c}
	\caption{Values of the adopted best (PES) parameters $ A_{2},A_{3},A_{4},A_{0} $ ( in KeV ) for the studied eight pairs of identical bands. $ N_{B} $ is the total number of bosons.}	\label{tab:4}\\
	\hline\noalign{\smallskip}
	&$ \;\;\;\; $  $ N_{B} $ $ \;\;\;\; $ &$ \;\;\;\; $  $ A_{2} $ $ \;\;\;\; $ &$ \;\;\;\; $   $ A_{3} $ $ \;\;\;\; $ &$ \;\;\;\; $   $ A_{4} $ $ \;\;\;\; $  &$ \;\;\;\; $   $ A_{0} $ $ \;\;\;\; $ \\
	
	\hline\noalign{\smallskip}
	\hline\noalign{\smallskip}
	
	
	
	$\;\; ^{162}Dy  \;\;\;\;$ &15 & -2.4667 & -0.5863 & 1.6665 & -0.3265  \\
	$\;\; ^{166}Er  \;\;\;\;$ & 15 & -1.6586 & -2.0341 & 4.4739 & -0.7875  \\
	\hline\noalign{\smallskip}
	$\;\; ^{162}Er  \;\;\;\;$ & 13 & -5.0526 & -2.5496 & 3.7667 & -0.9375  \\
	$\;\; ^{166}Yb  \;\;\;\;$ & 13 & -5.3088 & -3.1366 & 4.0554 & -0.925  \\
	\hline\noalign{\smallskip}
	$\;\; ^{162}Yb  \;\;\;\;$ & 11 & -4.84 & -1.6163 & 3.6775 & -0.9  \\
	$\;\; ^{166}Hf  \;\;\;\;$ & 11 & -2.8484 & -1.9547 & 3.9131 & -0.8625  \\
	\hline\noalign{\smallskip}
	$\;\; ^{160}Dy  \;\;\;\;$ & 14 & -1.9568 & -0.8838 & 1.1005 & -0.3  \\
	$\;\; ^{168}Yb  \;\;\;\;$ & 14 & -5.3088 & -3.1366 & 4.0554 & -0.925  \\
	\hline\noalign{\smallskip}
	$\;\; ^{160}Er  \;\;\;\;$ & 12 & -3.0403 & -2.3636 & 4.1401 & -0.8625  \\
	$\;\; ^{168}Hf  \;\;\;\;$ & 12 & -3.463 & -2.4694 & 4.039 & -0.875  \\
	\hline\noalign{\smallskip}
	$\;\; ^{158}Dy  \;\;\;\;$ & 13 & -1.6288 & -1.1822 & 1.0095 & -0.288  \\
	$\;\; ^{170}Hf  \;\;\;\;$ & 13 & -3.1845 & -3.395 & 4.497 & -0.8375  \\
	\hline\noalign{\smallskip}
	$\;\; ^{158}Er  \;\;\;\;$ & 11 & -1.6586 & -2.0541 & 4.4739 & -0.7875  \\
	$\;\; ^{170}W  \;\;\;\;$ & 11 & -0.9761 & -2.4841 & 4.7606 & -0.7546  \\
	\hline\noalign{\smallskip}
	$\;\; ^{156}Dy  \;\;\;\;$ & 12 & -1.5043 & -1.2135 & 0.9961 & -0.3  \\
	$\;\; ^{172}W  \;\;\;\;$ & 12 & -0.8852 & -1.4675 & 1.0599 & -0.313  \\
	
	\hline\noalign{\smallskip}
\end{longtable}

\section{Conclusion}

By using a novel three parameters collective rotational formula (CRF3), the positive parity ground state excitation energies are calculated for sixteen nuclei in rare-earth region. The optimized three parameters are deduced by using a computer simulated search program in order to obtain a minimum root mean square deviation of the calculated excitation energies from the measured ones. The potential energy surfaces are calculated by using the sd-version of the interacting boson model.

The problem of low-spin identical bands in normal deformed nuclei in rare-earth region is treated. We have exhibited identical bands in eight pairs of conjugate even-even nuclei of widely dispersed spanning as much as sixteen mass unit. Each pair with the same F-spin and projections $ \pm F_{0} $ values have identical product of valence proton and neutron numbers $ N_{p}N_{n} $ values. Also, the values of dynamical moments of inertia for each identical band pair are approximately the same. We extracted all the identical band symmetry parameters like P-factor, saturation parameter, and structure factor which all depend on $ N_{p} $ and $ N_{n} $. The pairing interaction energy, the quadrupole transition probabilities, and the energy ratios are also treated.

\bibliography{ref}

\end{document}